\pdfoutput=1  
\documentclass[journal,twoside,web]{ieeecolor2}
\usepackage{generic}
\usepackage{cite}
\usepackage{amsmath,amssymb,amsfonts}
\usepackage{algorithmic}
\usepackage{graphicx}
\usepackage{textcomp}
\def\BibTeX{{\rm B\kern-.05em{\sc i\kern-.025em b}\kern-.08em
    T\kern-.1667em\lower.7ex\hbox{E}\kern-.125emX}}
\begin{document}
\title{Cross-Anesthetic ECoG State Decoding Fails at the Decision Threshold, Not the Representation}

\author{Kunkun Zhang, Qianwei zhou
\thanks{Manuscript submitted 2026. This work was supported by the National Natural Science Foundation of China (Grant no.62271448.}%
\thanks{The authors are with Zhejiang University of Technology (e-mail: zkk@zjut.edu.cn, zqw@zjut.edu.cn). (\emph{Corresponding author: Qianwei Zhou})}}

\maketitle
\raggedbottom   

\begin{abstract}
Decoders of anesthetic state from cortical activity fail across drug classes, most notoriously ketamine, but reported accuracy cannot say whether the neural representation or only the decision threshold has failed; we separate the two in a controlled preparation with ground-truth labels. We decoded awake versus anesthetized from mouse electrocorticography (the 250-Hz-bandlimited local field potential, sampled at 1875 Hz) under leave-one-anesthetic-out evaluation across five mechanistically distinct anesthetics (isoflurane, dexmedetomidine, ketamine, propofol, and midazolam), comparing a spatially blind band-power decoder, covariance/Riemannian representations, and Riemannian domain adaptation, with all statistics at the session level and mouse-level cluster bootstrapping for the ketamine fold. The representation transfers: band-power ranks awake versus anesthetized at a session AUROC of at least 0.96 on every held-out drug, ketamine included (0.980, cluster confidence interval 0.821 to 1.000). The failure is confined to the threshold: across three representations the ketamine ranking is near-invariant while its balanced accuracy swings from chance to high, and a permutation test is significant for ranking (p = 0.0025) but not for fixed-threshold accuracy (p = 0.3795). Riemannian domain adaptation is net-negative. A causal, label-free threshold anchored to the subject's own pre-induction baseline fixes ketamine (balanced accuracy 0.50 to 0.85) and dominates domain adaptation. Because the ketamine test sessions come from three mice that also contribute training drugs, this is within-subject cross-drug transfer; we do not claim population-level transfer across subjects. In cross-drug state decoding the actionable failure is calibration, not representation.
\end{abstract}

\begin{IEEEkeywords}
Anesthesia depth monitoring, brain--computer interfaces, calibration, domain adaptation, electrocorticography, transfer learning.
\end{IEEEkeywords}

\section{Introduction}
\label{sec:introduction}
\IEEEPARstart{A}{ decoder} trained to tell the anesthetized brain from the awake one, then given an anesthetic it has never seen, can fail in two ways a single accuracy figure cannot distinguish: it may lose the neural distinction, or keep it intact and merely place the decision boundary wrongly. The two failures call for opposite remedies, yet accuracy fuses them. Decoding awake versus anesthetized from mouse electrocorticography (ECoG) and holding out ketamine, a channel-pooled band-power model with no spatial information still ranks unconscious ketamine recordings above awake ones at a session-level AUROC of 0.98; yet, using the boundary inherited from the other four anesthetics, it labels every anesthetized ketamine session awake. Depth-of-anesthesia indices were validated largely on GABAergic hypnotics and are known to read paradoxically for agents with a different cortical signature, ketamine among them \cite{dahaba2005,purdon2015}. We show that here the failure is a miscalibrated threshold, not a missing representation, and that a simple, deployable correction repairs it where the field's standard tool does not.

The five anesthetics we study span four mechanisms: isoflurane (an inhalational ether), dexmedetomidine (an $\alpha_2$-adrenergic agonist), ketamine (an NMDA-receptor antagonist), and propofol and midazolam (both GABA-A). Their spectral signatures diverge in ways a band-power model can see: propofol and isoflurane produce a sustained delta-dominated slowing, dexmedetomidine produces sleep-like spindles, midazolam adds prominent beta, and ketamine mixes gamma with slow-delta rather than holding the sustained delta-dominance of the others \cite{purdon2013,purdon2015,akeju2014,akeju2016}. That contrast is the crux: ketamine's slow-delta still projects onto the source-trained discriminant, so the ranking survives, while its high-frequency content shifts the whole ketamine distribution below the inherited boundary. A band ablation confirms this high-frequency content is discriminative: extending the analysis to 200~Hz raises ketamine AUROC from 0.767 to 0.867. Holding out a drug thus removes a mechanism from training, strictly so for dexmedetomidine and ketamine, whose targets are shared by no other agent, but only partly for propofol and midazolam, which share GABA-A (propofol, whose partner remains in training, is accordingly our easiest fold). The generator of the shift is explicit and biological, and this is a diagnosis the clinic cannot perform on itself: it has no independent ground-truth readout of consciousness, cannot hold the recording montage and pipeline fixed across drug classes, and cannot ethically withhold a drug class in order to test on it. A controlled preparation supplies exactly that ground truth and fixed pipeline.

Machine-learning decoders of anesthetic state are built almost exclusively from human frontal scalp EEG, classify sedation level within a single drug or from a drug-pooled training set, and are judged by discrimination alone---AUROC, accuracy, or F1 \cite{shalbaf2013,gu2019,saadeh2019,nagaraj2018,ramaswamy2019,ramaswamy2020,wang2022}. The ``drug-independent'' estimators pool drugs and cross-validate across subjects rather than holding a drug out; the only demonstrated cross-drug transfer is between agents that share a GABAergic action and a common slow-wave signature (propofol and sevoflurane), and it explicitly excludes ketamine \cite{abel2021}; and where ketamine does appear, the reported remedy is a drug-specific model rather than a drug-invariant one \cite{kashkooli2020}. Calibration is essentially never reported, and no study performs strict leave-one-drug-out onto a mechanistically distinct held-out anesthetic, let alone on ECoG. When transfer degrades, the reflex is to treat it as a representation problem: richer features, self-supervised pretraining, or domain adaptation, with covariance-based Riemannian alignment and its Euclidean variant the de facto standard tools \cite{barachant2012,yger2017,zanini2018,hewu2020}. Yet the accuracy that motivates this reflex fuses two quantities a ranking metric keeps apart: whether the representation still orders brain states correctly under shift, and whether the inherited boundary still sits in the right place \cite{fawcett2006,bradley1997,steyerberg2010}. What is genuinely unknown is not that a threshold can drift under distribution shift, which is expected, but where, in a real biological transfer with a known causal generator, the failure lives; whether the standard remedy addresses it; and whether the domain itself supplies a cheaper, correct alternative.

Here we show, in this preparation, that the representation transfers while the decision boundary does not: the band-power decoder ranks every held-out drug well, yet at the threshold it inherits it calls the unconscious ketamine sessions awake. The correction is to re-anchor the threshold, not to enrich or realign the representation; we introduce no new estimator, and the contribution is the diagnosis. The ketamine result rests on five anesthetized test sessions, a ceiling we state throughout. Our contributions are as follows.

\begin{itemize}
\item Under leave-one-anesthetic-out across five anesthetics spanning four mechanisms, a spatially blind band-power decoder ranks awake versus anesthetized on every held-out drug, ketamine included (session AUROC 0.98, 95\% CI [0.885, 1.000] session-level and [0.821, 1.000] mouse-cluster; a lower per-epoch value of 0.68 that the session figure improves on by temporal averaging). Because that ranking holds while balanced accuracy at the inherited 0.5 threshold sits at chance, cross-drug transfer fails at the decision boundary, not the representation. Because the ketamine test sessions come from three mice that also contribute training drugs, this is within-subject cross-drug transfer; we do not claim population-level transfer across subjects.
\item The localization is constructive, not a contrast of two metrics: across the three representations the ketamine ranking is near-invariant (0.98/0.94/0.96, the richer covariance representation ranking ketamine slightly lower) while its balanced accuracy swings from chance to high, and the same baseline information corrects ketamine at the threshold but not at the feature level. A permutation test corroborates.
\item The field's standard remedy is aimed at the wrong quantity: Riemannian domain adaptation, evaluated as a non-causal upper bound, is net-negative, buying one drug at another's expense, whereas a baseline-anchored threshold does better without changing the decoder.
\item The correction is deployable: no anesthetized labels, no new model, no change to clinical workflow, and a validity domain that is characterized rather than assumed.
\end{itemize}

\section{Methods}

\subsection{Animals, Recordings, and the Signal Band}
All animal procedures were approved by the host institution's Institutional Animal Care and Use Committee (IACUC protocol no.~\textit{XXXX}) 
and conducted in accordance with relevant guidelines. Electrocorticography (ECoG) was recorded from mice with an Intan RHS system (one file per channel; amplifier least-significant bit 0.195~$\mu$V). The neural signal is the Intan low/LFP band: a 250~Hz Bessel low-pass with an on-chip downsample factor of 16, so the amplifier band is sampled at $30\,000/16 = 1875$~Hz rather than the 30~kHz of the raw timebase and digital lines (verified on all folders by a sample-count ratio of 16). Analyses therefore concern the $\leq 250$~Hz LFP band, and we make no claim about wideband high-gamma above 250~Hz.

Channel counts vary from 7 to 32, and vary \emph{within} a mouse (channels enabled or disabled across sessions, not remapped). The global channel intersection across mice is only four, so no shared cross-mouse montage exists; decoders that need a fixed channel set are trained \emph{per mouse} on that mouse's common montage. Three well-sampled ``core'' mice supply montages of 15, 16, and 19 channels; a fourth mouse supplies only isoflurane and dexmedetomidine and is excluded from the per-mouse decoders. The full labeled set is 129 sessions (dexmedetomidine 36, isoflurane 39, ketamine 16, midazolam 19, propofol 19).

\subsection{Anesthetics and Phase Labels}
Five anesthetics span four mechanisms: isoflurane (inhalational ether), dexmedetomidine ($\alpha_2$-adrenergic agonist), ketamine (NMDA-receptor antagonist), and propofol and midazolam (both GABA-A). Each recording is labeled by protocol phase: \emph{base} (awake, pre-drug) and \emph{post} (awake after return of consciousness, protocol-confirmed) are class~0 (awake); \emph{effect} (drug on board, unconscious) is class~1 (anesthetized); \emph{recover} (emergence) is transitional and excluded. Labels are carried from the paired imaging session's phase timing via a per-recording offset; multiple manifest rows can share one recording, each cutting its own time window.

\subsection{Preprocessing and Epoching}
The 1875~Hz LFP is decimated to 500~Hz (Nyquist 250~Hz), notch-filtered at 60/120/180~Hz ($Q=30$), and band-passed 0.5--200~Hz (4th-order, upper edge auto-clipped below Nyquist). Each session is z-scored per session, making features approximately relative band fractions. Epochs are 2~s windows at a 1~s step (50\% overlap). Because effect recordings are ${\sim}10\times$ longer than awake ones, each session is capped at 186 epochs by uniform sub-sampling; this jointly fixes the ${\sim}92/8$ class imbalance and long-session dominance.

\subsection{Decoders}
Three decoders separate \emph{representation} from \emph{alignment}.
\begin{itemize}
\item \textbf{BP}---channel-pooled log band-power in eight bands (1--4, 4--8, 8--13, 13--30, 30--55, 65--95, 105--145, 155--200~Hz), averaged over channels, classified by $L_2$-regularized logistic regression with balanced class weights. Eight dimensions, \emph{no spatial information}.
\item \textbf{RN}---per-mouse montage OAS covariance $\rightarrow$ Riemannian tangent space $\rightarrow$ logistic regression. Spatial, \emph{no} domain alignment.
\item \textbf{RA}---RN with each training domain recentered to the identity (whitening by the domain's Riemannian mean), and the held-out drug recentered by its own pooled mean at test. This uses the test drug's data to set the recentering and is applied offline; we label RA a \emph{non-causal upper bound} wherever it appears.
\end{itemize}
RN and RA require same-dimension covariances, so the BP/RN/RA comparison runs per mouse on the three core mice with test predictions pooled (configuration~C1). BP is channel-pooled and montage-independent; where a non-degenerate statistical test is needed it is also run on all mice (configuration~C2).

\subsection{Baseline-Anchored Threshold Calibration}
The deployable correction changes only the decision threshold. For a held-out drug, the threshold is set to the 95th percentile of BP scores on that drug's own base (pre-induction, awake) sessions, then applied to the test scores. It uses no anesthetized labels (the identity of baseline epochs is protocol knowledge), it is causal (the baseline precedes induction), and $q = 0.95$ was fixed a priori; we report the full $q$-sweep. We contrast it with feature-level baseline normalization---pairing each session to its matched base and normalizing log band-power by that baseline (a log-ratio, ``SUB''), applied to train and test with the model retrained at a fixed 0.5 threshold---which uses the same information at a different point.

\subsection{Evaluation and Statistics}
The primary protocol is leave-one-anesthetic-out (LOAO): train on four drugs, test on the held-out fifth. The secondary protocol is leave-one-mouse-out (LOMO). Grouping is by session. Because 2~s epochs overlap 50\% and are autocorrelated, the effective sample size is far below the nominal epoch count, so \emph{all inferential statistics are computed at the session level}; epoch-level AUROC is reported only descriptively. Confidence intervals are session-resampled bootstraps (2000 iterations); permutation tests shuffle session labels (2000 iterations). We report AUROC, AUPRC, and balanced accuracy, and never raw accuracy. Because sessions within a mouse share montage, placement, and physiology, for the ketamine fold we additionally resample mice first (with replacement) and then sessions within each sampled mouse (a mouse-level cluster bootstrap). In configuration~C1 the BP decoder assigns 100\% of ketamine effect sessions to ``awake,'' so its balanced accuracy is identically 0.500, its bootstrap CI is a point mass, and its permutation null is degenerate; the non-degenerate ketamine claim therefore comes from the all-mice run (C2).

\section{Results}

\subsection{The Representation Transfers Across Drugs}
Under LOAO, the spatially blind band-power decoder ranks awake versus anesthetized at session AUROC $\geq 0.96$ on every held-out drug (C1): dexmedetomidine 0.980, isoflurane 0.990, ketamine 0.980, midazolam 0.960, propofol 1.000. For ketamine the session AUROC is 0.980, 95\% CI [0.885, 1.000], from 5 effect / 15 total test sessions across three mice.

The transfer is partial and session-level: the ketamine epoch-level AUROC is 0.680, well above chance but far below the session figure, which averages up to 186 epochs. No epoch-level CI is available and epochs are autocorrelated, so we treat 0.680 as descriptive. Resampling mice then sessions widens the ketamine session-AUROC CI to [0.821, 1.000], from the session-only [0.885, 1.000]; the lower bound falls ${\sim}0.06$ but remains far from chance, so the ranking claim survives clustering. The LOAO mean session AUROC is 0.982, cluster CI [0.950, 1.000]. All 15 ketamine test sessions come from the three core mice, which also supply the four training drugs, so this is \emph{within-subject cross-drug} transfer; population-level transfer to a novel subject on ketamine is untested (see Discussion).

\subsection{The Failure Localizes to the Decision Threshold}
\begin{figure}[!t]
\centerline{\includegraphics[width=\columnwidth]{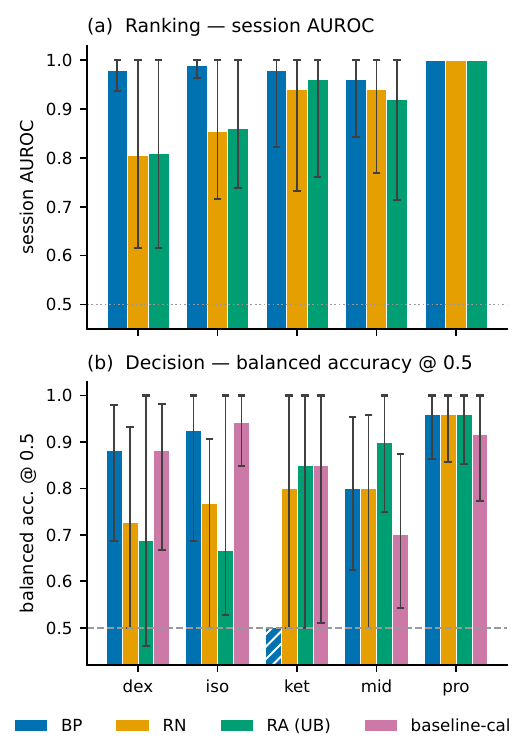}}
\caption{Ranking transfers; the threshold breaks. Leave-one-anesthetic-out, five held-out drugs, three decoders: band-power (BP, 8-dim, spatially blind), covariance-tangent (RN), and Riemannian domain adaptation (RA, a non-causal upper bound). (a) Session AUROC---band-power ranks every drug $\geq 0.96$, while the covariance methods lose ranking on dexmedetomidine and isoflurane, so a richer representation does not help. (b) Balanced accuracy at the inherited 0.5 threshold, with baseline calibration added---ketamine collapses to 0.500 under band-power (degenerate: 100\% of effect sessions called awake) and recovers to ${\sim}0.85$ under RN/RA/calibration; other drugs stay high. Error bars are mouse-cluster 95\% CIs (resample mice, then sessions within mouse) on \emph{every} bar; the ketamine band-power bar in (b) is degenerate (all sessions called awake, a point-mass CI) and is hatched without a whisker. Dashed line: chance. Ketamine $n_{\mathrm{eff}} = 5$ (15 total test sessions, three mice). Colors denote methods consistently across all figures: band-power (BP) blue, covariance-tangent (RN) orange, Riemannian adaptation (RA) green, baseline-calibration purple; Fig.~\ref{fig:calib} adds fixed-0.5 gray and oracle vermillion.}
\label{fig:core}
\end{figure}

Across three very different representations the ketamine ranking is nearly invariant while the fixed-threshold balanced accuracy swings wildly (Table~\ref{tab:localize}). The richer covariance representation (RN) ranks ketamine \emph{lower} than band-power (0.940 $<$ 0.980), so a better representation does not ``fix'' ketamine; it merely lands the scores on both sides of 0.5. Figure~\ref{fig:core} shows the pattern across all five drugs: a near-flat band of session AUROC over the top, and a jagged balanced accuracy below in which ketamine collapses to 0.500 under band-power and recovers under RN, RA, and calibration.

The non-degenerate test uses all mice (C2): ketamine AUROC 0.933 (permutation $p = 0.0025$, significant) but balanced accuracy at 0.5 equal to 0.583 ($p = 0.3795$, not significant)---same model, same fold, one metric significant and one not. The failure is the threshold, not the ranking.

\begin{table}[!t]
\caption{Ketamine ranking is representation-invariant; only the threshold moves (configuration C1).}
\label{tab:localize}
\centering
\setlength{\tabcolsep}{4pt}
\begin{tabular}{lcc}
\hline
Ketamine (LOAO) & session AUROC & session bal.\ acc.\ @0.5 \\
\hline
Band-power (8-dim, pooled) & 0.980 & 0.500\textsuperscript{$\dagger$} \\
Covariance $\rightarrow$ tangent (RN) & 0.940 & 0.800 \\
RN + recentering (RA, UB) & 0.960 & 0.850 \\
\hline
\multicolumn{3}{p{0.92\columnwidth}}{\footnotesize \textsuperscript{$\dagger$}Degenerate: band-power calls 100\% of ketamine effect sessions awake, so balanced accuracy is identically 0.500 with a point-mass CI. UB: non-causal upper bound.}
\end{tabular}
\end{table}

\subsection{Riemannian Domain Adaptation Is Net-Negative}
As an offline, non-causal upper bound, RA does not improve on average (C1): mean session AUROC 0.910 (RA) $<$ 0.982 (BP); mean session balanced accuracy 0.8127 (RA) $<$ 0.8132 (doing nothing); mean epoch AUROC 0.812 $<$ 0.839. RA buys ketamine (balanced accuracy 0.500 $\rightarrow$ 0.850) only by damaging dexmedetomidine and isoflurane (0.883 $\rightarrow$ 0.688; 0.925 $\rightarrow$ 0.667). The damage is present already at RN, i.e.\ in the representation, before any recentering (dexmedetomidine AUROC 0.980 $\rightarrow$ 0.806; isoflurane 0.990 $\rightarrow$ 0.853). Since RA is an upper bound, its net-negativity is conservative: even best-case realignment loses on average.

\subsection{Baseline-Anchored Calibration Fixes Ketamine and Dominates RA}
\begin{figure}[!t]
\centerline{\includegraphics[width=\columnwidth]{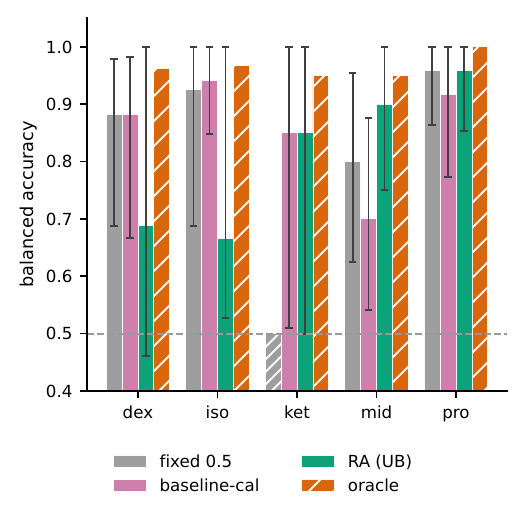}}
\caption{A deployable fix that dominates domain adaptation. Per-fold balanced accuracy for the fixed-0.5 threshold, baseline-anchored calibration ($q{=}0.95$), RA (non-causal upper bound), and the oracle ceiling. Baseline calibration fixes ketamine (0.500 $\rightarrow$ 0.850, matching the oracle direction) while leaving dexmedetomidine/isoflurane/propofol untouched, whereas RA buys ketamine only by damaging dexmedetomidine and isoflurane. All bars carry mouse-cluster 95\% CIs (the same method as Fig.~\ref{fig:core}); the ketamine baseline-cal CI is [0.510, 1.000]---wide, touching chance, reflecting the three-mouse / five-effect-session ceiling. The oracle is a ceiling (no CI), and the ketamine fixed-0.5 bar is degenerate (hatched, no whisker). Colors follow the shared mapping of Fig.~\ref{fig:core}: fixed-0.5 gray, baseline-calibration purple, RA green, oracle vermillion. Dashed line: chance.}
\label{fig:calib}
\end{figure}

Thresholding BP scores at the 95th percentile of the target drug's own base sessions (Table~\ref{tab:calib}, Fig.~\ref{fig:calib}) is the \emph{only} intervention that raises the mean balanced accuracy (0.813 $\rightarrow$ 0.858); it fixes ketamine (0.500 $\rightarrow$ 0.850, effect-called-awake 1.00 $\rightarrow$ 0.20) while leaving dexmedetomidine, isoflurane, and propofol essentially untouched, with no representation change and no domain adaptation. It \emph{dominates} RA: the same ketamine fix without RA's damage. The fix is robust to $q$ (ketamine balanced accuracy 0.900 at $q{=}0.90$, 0.850 at $q{=}0.95$ and $0.99$, versus 0.500 fixed).

We report the power on the fix honestly. Under the mouse-cluster bootstrap the ketamine baseline-cal balanced accuracy is 0.850 but with a wide CI [0.510, 1.000] (three mice, five effect sessions). The fix is a clear point improvement over fixed-0.5 (0.500) that matches the oracle direction, but it is not a tightly bounded estimate---the ketamine power ceiling applies to the correction as much as to the failure.

\begin{table}[!t]
\caption{Balanced accuracy of the deployable fix versus alternatives (configuration C1). Baseline calibration fixes ketamine and dominates Riemannian adaptation.}
\label{tab:calib}
\centering
\setlength{\tabcolsep}{4pt}
\begin{tabular}{lccccc}
\hline
Drug & $n_{\mathrm{eff}}$ & fixed 0.5 & baseline-cal & RA (UB) & oracle \\
\hline
Ketamine & 5 & 0.500 & \textbf{0.850} & 0.850 & 0.950 \\
Dexmedetom. & 13 & 0.883 & 0.883 & 0.688 & 0.962 \\
Isoflurane & 15 & 0.925 & 0.942 & 0.667 & 0.967 \\
Midazolam & 5 & 0.800 & 0.700 & 0.900 & 0.950 \\
Propofol & 6 & 0.958 & 0.917 & 0.958 & 1.000 \\
\hline
Mean & & 0.8132 & \textbf{0.8582} & 0.8127 & 0.9656 \\
\hline
\multicolumn{6}{p{0.95\columnwidth}}{\footnotesize UB: non-causal upper bound. Oracle: a ceiling using the held drug's own labels. Ketamine baseline-cal mouse-cluster CI [0.510, 1.000] (Fig.~\ref{fig:calib}).}
\end{tabular}
\end{table}

\subsection{Feature-Normalization Control}
Applying the \emph{same} baseline information at the feature level and retraining at fixed 0.5 (Table~\ref{tab:featnorm}) is not a strawman: it genuinely improves dexmedetomidine (balanced accuracy 0.883 $\rightarrow$ 0.974, AUROC $\rightarrow$ 1.000) and propofol (0.958 $\rightarrow$ 1.000). Yet it does exactly nothing for ketamine (balanced accuracy 0.500, effect-called-awake 1.00, identical to no normalization) and even lowers ketamine AUROC (0.980 $\rightarrow$ 0.900), because retraining relearns a source-drug boundary still misplaced for the held-out drug. Same information, different application point: only re-anchoring the decision boundary works. (Per-session z-scoring already makes features relative band fractions; cross-recording baseline normalization is a distinct operation.)

\begin{table}[!t]
\caption{Same baseline information, different application point (configuration C1). Only threshold anchoring corrects ketamine.}
\label{tab:featnorm}
\centering
\setlength{\tabcolsep}{3.5pt}
\begin{tabular}{lccccc}
\hline
 & ket bacc & ket AUROC & dex bacc & pro bacc & mean bacc \\
\hline
BP, fixed 0.5 & 0.500 & 0.980 & 0.883 & 0.958 & 0.813 \\
BP, baseline threshold & \textbf{0.850} & 0.980 & 0.883 & 0.917 & \textbf{0.858} \\
Feature-norm, retrain & 0.500 & 0.900 & \textbf{0.974} & \textbf{1.000} & 0.825 \\
\hline
\end{tabular}
\end{table}

\subsection{Validity Domain}
Baseline calibration assumes that the awake baseline matches the awake recovery (base $\approx$ post). Base-vs-post separability measures how violated that is, and rank-predicts the calibration gain across all five drugs (Table~\ref{tab:validity}): Spearman $\rho = -1.000$ (exact-permutation $p = 0.0167$; $n = 5$ drugs---suggestive, not established). Calibration helps most where base $\approx$ post (ketamine) and hurts where they differ (midazolam).

\begin{table}[!t]
\caption{Base-vs-post separability rank-predicts where calibration helps.}
\label{tab:validity}
\centering
\setlength{\tabcolsep}{6pt}
\begin{tabular}{lcc}
\hline
Drug & base-vs-post separability & baseline-cal bacc gain \\
\hline
Ketamine & 0.500 & $+0.350$ \\
Isoflurane & 0.567 & $+0.017$ \\
Dexmedetom. & 0.623 & $0.000$ \\
Propofol & 0.690 & $-0.042$ \\
Midazolam & 0.933 & $-0.100$ \\
\hline
\end{tabular}
\end{table}

\subsection{Two Opposite Dissociations, One Root Cause}
The \emph{post} condition (drug on board, behaviorally recovered) is a dissociation most anesthesia studies lack, and it exposes two opposite errors that share one cause---the decoder reads pharmacology, not consciousness. Ketamine-effect (unconscious, drug on board) is called awake (100\% under BP in C1; 5/6 in C2); midazolam-post (conscious, drug residual) is called anesthetized. Post-midazolam is the diagnostic case: read anesthetized in 80\% of sessions under BP, halved to 40\% by Riemannian adaptation but not eliminated, and worsened to 100\% by baseline calibration. A method that \emph{halves} it and a method that \emph{worsens} it both leave a residual, which is the signature of a genuine pharmacological confound (benzodiazepine $\beta$ persisting after behavioral recovery), not a threshold error.

\begin{figure}[!t]
\centerline{\includegraphics[width=\columnwidth]{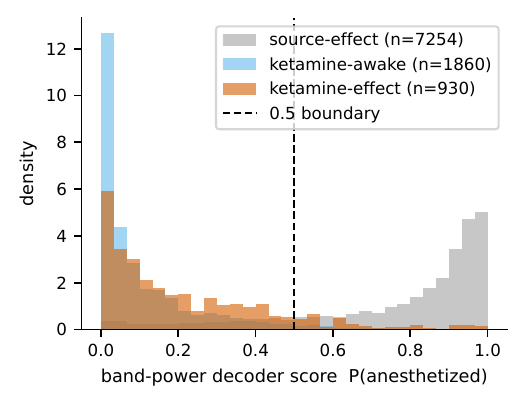}}
\caption{Epoch-score shift, not a resolvable split. Per-epoch band-power scores for the ketamine fold: ketamine-effect (mean 0.215) sits above ketamine-awake (0.107)---ranking preserved---but far below source-effect (0.762), with 89\% of ketamine-effect epochs below 0.5. The ketamine-effect distribution is unimodal (Hartigan dip $p = 0.97$; also at 0.5~s windows, $p = 0.99$), so the shift is not a resolvable mix of awake-like and anesthetized-like epochs. $n = 930/1860/7254$ epochs (ketamine-effect / ketamine-awake / source-effect).}
\label{fig:epoch}
\end{figure}

Figure~\ref{fig:epoch} shows the ketamine shift at epoch resolution: ketamine-effect scores (mean 0.215) sit above ketamine-awake (0.107)---ranking preserved---but far below source-effect (0.762), and 89\% of ketamine-effect epochs fall below 0.5. The ketamine-effect distribution is unimodal (Hartigan dip $p = 0.97$; also unimodal at 0.5~s windows, $p = 0.99$); a bimodality coefficient and a two-component mixture flag ``two modes'' for all three distributions, including the plainly unimodal source-effect, so they are artifacts of skew rather than evidence of bimodality. One account of the shift---that the ketamine gamma/slow-delta alternation is sub-window and averages within epochs---is consistent with the data but not resolved by it, and we do not claim it. The high-frequency content is nonetheless discriminative: extending the bands to 200~Hz lifts ketamine AUROC 0.767 $\rightarrow$ 0.867.

\subsection{Two Negative Results}
\emph{Causal streaming is structurally infeasible at this timescale.} On 39 loss/return-of-consciousness trajectories (ketamine $n = 5$, under-powered), epoch AUROC is: a band-power floor 0.609; non-causal offline Riemannian recentering 0.620 (an upper bound ${\approx}$ the floor); naive causal online 0.530; and memory-augmented online test-time adaptation 0.559. The running Riemannian mean converges at ${\sim}1250$~s---approximately the recording length---so the estimator never warms up (offline 0.793 vs.\ causal 0.479). Even a perfect online method cannot beat band-power here.

\emph{Zero-shot drug-presence invariance is a transfer gap, not entanglement.} Projecting out the drug-presence (base-vs-post) direction learned from \emph{training} drugs does nothing to the held-out drug's confound (post-midazolam 0.40 unchanged for $k = 1\ldots20$ removed directions) and does not destroy state (AUROC ${\sim}0.91$). The oracle direction (the held drug's own base/post) removes it fully (0.40 $\rightarrow$ 0.00) without hurting state (midazolam AUROC 1.000). Each drug's on-board signature is its own direction; this is a few-shot transfer problem, not fundamental entanglement. Consistent with this, \emph{post} sessions carry a weak drug trace---multiclass drug-identity accuracy 0.353 versus 0.20 chance---so the on-board signature is drug-specific rather than a single shared axis.

\section{Discussion}

\subsection{What the Result Means}
Cross-drug transfer failure in this setting is not a representation failure but a calibration failure, and the two are separable: a ranking metric (AUROC) is invariant to where the boundary sits, while thresholded accuracy is not, so a decoder can rank a held-out drug almost perfectly yet mislabel all of it. The field's reflex---richer representations, self-supervised pretraining, covariance/Riemannian domain adaptation---targets the representation, which is the quantity that did \emph{not} fail here; accordingly the standard tool (RA) is net-negative on this transfer, trading one drug for another. The actionable message is to report and correct calibration, not to enrich the representation, when transfer degrades under a prior or covariate shift with a known generator.

\subsection{Clinical Parallel (an Analogy)}
The ketamine failure echoes a clinical observation: processed-EEG depth indices, developed largely on GABAergic hypnotics, can rise paradoxically when ketamine is present, because ketamine's cortical signature (increased high-frequency power, less sustained delta) departs from the slow-wave pattern the indices were calibrated on. This is documented for adjunctive, sub-anaesthetic ketamine added to a GABAergic agent---BIS rose 33 $\rightarrow$ 46 under sevoflurane \cite{hans2005}, higher BIS at hypnotic endpoints under propofol \cite{sakai1999}, 44 $\rightarrow$ 59 added to propofol--fentanyl \cite{hirota1999}, catalogued as a paradoxical BIS increase \cite{dahaba2005}. This is an analogy, not an equivalence: our decoder is band-power with logistic regression, not the proprietary BIS algorithm; the clinical effect is modest (the index rises to ${\sim}46$--59, not to awake levels) and dose-dependent; and our effect is larger and cleaner (100\% of effect sessions mislabeled). We therefore keep the clinical parallel in the Discussion and rest no result on it.

\subsection{Pharmacology, Not Consciousness}
The two opposite dissociations make one point: a zero-shot cross-drug decoder tracks each drug's pharmacological signature, not the level of consciousness. Ketamine-effect is called awake because its spectrum lacks sustained delta-dominance. This interpretation is consistent with independent evidence obtained through a different modality and analysis: using wide-field calcium imaging and neuronal-avalanche statistics, surgical-plane ketamine produces cortical dynamics that differ qualitatively from other anesthetics, and a subset of surgical-plane ketamine recordings are statistically indistinguishable from quiet wakefulness \cite{curic2024}. That an entirely separate measurement can render ketamine anesthesia awake-like offers cross-modal, cross-method support for our finding that a source-trained decoder ranks ketamine-anesthetized activity as awake---while underscoring that these are convergent proxies, not a shared measurement of consciousness. Midazolam-post is called anesthetized because benzodiazepine $\beta$ persists after behavioral recovery. Post-midazolam is therefore a genuine pharmacological confound, not miscalibration: read anesthetized in 80\% of sessions under BP, halved to 40\% by Riemannian adaptation but not eliminated, and worsened to 100\% by baseline calibration. That a domain-adaptation method only halves it and the baseline fix worsens it is the signature of a real pharmacological trace, not a threshold any method could re-place, and for a \emph{state} decoder it should not be corrected away. Importantly, the invariance analysis shows the consciousness signal \emph{is} present and few-shot recoverable given a little target-domain data, so ``reads pharmacology'' is a statement about the zero-shot decoder, not about the ceiling of the signal.

\subsection{Why Baseline Anchoring Is the Right Primitive}
The correct fix rescales the decision to the drug's own attenuated score axis using information that always exists in the operating room---a pre-induction awake recording---and no anesthetized labels. Its validity domain is predictable from first principles: it helps when the subject's awake baseline matches its awake recovery and hurts when they differ, and base-vs-post separability rank-predicts the gain ($\rho = -1.000$, $n = 5$, suggestive). Midazolam is the failure case (largest base-vs-post separability, 0.933; calibration gain $-0.100$), and it fails for a principled reason: the post-drug awake state is spectrally distinct from the pre-drug awake state, so a baseline anchor mis-sets the threshold. A separating threshold exists for midazolam (oracle balanced accuracy 0.950) but requires \emph{post}, which is unavailable prospectively---the structural limit of the method.

\subsection{Limitations}
Ketamine is under-powered: five anesthetized (effect) test sessions, 15 total, from three mice; the mouse-cluster bootstrap widens the ketamine AUROC CI to [0.821, 1.000] (from [0.885, 1.000])---still above chance, but the hard ceiling on the headline, and no more ketamine data exists. Because the ketamine test sessions come from three mice that also contribute training drugs, this is within-subject cross-drug transfer; we do not claim population-level transfer across subjects (this also means the ketamine result is not confounded by subject novelty). LOMO is comparatively easy---per-mouse session AUROC 0.971 / 0.938 / 0.924 (a fourth mouse degenerate, excluded)---because every drug is present in training: subject is the easy axis, drug the hard one. The analysis is limited to the $\leq 250$~Hz LFP band, so the ${>}250$~Hz gamma that most distinguishes ketamine clinically is unavailable, and while the 80--200~Hz ablation (0.767 $\rightarrow$ 0.867) shows more high-frequency content helps, it does not close the gap. RA is a non-causal upper bound throughout; its net-negativity is conservative, but its positive numbers are not achievable online. Riemannian montages are pooled across three core mice. The validity-domain law rests on $n = 5$ drugs ($\rho = -1.000$): suggestive, not established. The streaming result rests on two full and five emergence-only ketamine trajectories. The ketamine epoch-score distribution is unimodal at both 2~s and 0.5~s, so the sub-window-alternation account is an interpretation we cannot confirm.

\subsection{Future Work}
Three pointers follow. First, cross-recording priors for the cold-start problem: causal streaming fails because the online Riemannian mean needs approximately one recording to converge, and a subject- or drug-level prior mean could warm-start it. Second, wideband re-acquisition (${>}250$~Hz) to test the ketamine high-frequency mechanism directly rather than inferring it from the $\leq 200$~Hz ablation. Third, more ketamine, and a novel-subject ketamine cohort, to move from within-subject to population-level cross-drug claims.

\section*{Acknowledgment}
The authors thank [colleagues/facilities].

\section*{Data and Code Availability}
The analysis code, the figure-generating scripts, and the frozen derived-result tables (CSV) that support the findings of this study can be made available to qualified researchers. 
Access to the raw electrocorticography recordings is subject to the host institution's data-sharing policy and is available from the corresponding author on reasonable request. 

\bibliographystyle{IEEEtran}
\bibliography{reference}


\end{document}